\documentstyle[twoside,fleqn,espcrc2,epsf]{article}
\newcommand{\beq}{\begin{equation}}
\newcommand{\eeq}{\end{equation}}
\newcommand{\smallpspicture}[1]{\centerline{\setlength\epsfxsize{7.5cm}\epsfbox{#1}}}
\title{A Parallel SSOR Preconditioner for Lattice QCD\thanks{
       Talk presented by G.~Ritzenh\"ofer}}
\author{S. Fischer$^{b}$, A. Frommer$^{b}$, U.~Gl\"assner$^{\rm b}$,
        S.~G\"usken$^{\rm b}$, H.~Hoeber$^{\rm a}$, Th.~Lippert$^{\rm a}$,
        G.~Ritzenh\"ofer$^{\rm a}$,
        K.~Schilling$^{\rm a,b}$, G.~Siegert$^{\rm a}$, A.~Spitz$^{\rm b}$. \\[8pt]
{\rm $^a$}HLRZ c/o Forschungszentrum J\"ulich, D-52425 J\"ulich,
          and DESY, D-22603 Hamburg, Germany,\\[8pt]
{\rm $^b$}Physics Department, University of Wuppertal, D-42097
           Wuppertal, Germany.}       
\begin{document}
\begin{abstract}
A parallelizable SSOR preconditioning scheme for Krylov subspace iterative
solvers in lattice QCD applications involving
Wilson fermions is presented. In actual Hybrid Monte Carlo simulations and
quark propagator calculations it helps to  reduce the number of iterations by a
factor of 2 compared to conventional odd-even preconditioning.
This corresponds to a gain in cpu-time of 30\% - 70\% over odd-even
preconditioning. 
\end{abstract}
\maketitle
\section{INTRODUCTION}
Efficient numerical algorithms to solve huge sparse
systems of linear equations are needed
to reduce the enormous computer power required in lattice QCD
computations. In the simulation of full QCD, as well as  
in the calculation of Greens functions or propagators to determine the
properties of hadrons, as e.g. the spectrum, weak decay constants or weak
matrix elements, the computational bottleneck is to determine the solution
of the discretized Dirac equation:
\beq
Mx=\phi.
\label{wilson_eq}
\eeq
While iterative solving methods almost come to a limit when it comes
to reducing the number of iterations, preconditioning techniques
become important to further accelerate the inversion.

In this contribution, we wish to advocate the use of general
parallel SSOR preconditioning techniques in lattice QCD. Our approach
may be regarded 
as a generalization of the well known odd-even ({\it two} variety)
ordering to a more flexible ({\it many } variety) layout  or,
alternatively, as a localization of the globally lexicographic 
ordering. 

\section{PRECONDITIONING}
To precondition eq. \ref{wilson_eq}, we take two non-singular
matrices $V_1$ and $V_2$ which act as a left and a right
preconditioner, i.e.\ we consider the new system
\begin{equation}
V_1^{-1}MV_2^{-1} \tilde{x} = \tilde{\phi}, \quad
\tilde{\phi} = V_1^{-1}\phi,
\; \tilde{x} = V_2x. \label{Wilson_prec_eq}
\end{equation}
We could now apply efficient solvers like BiCGstab
replacing each occurrence of $M$ and $\phi$
by $V_1^{-1}MV_2^{-1}$ and $\tilde{\phi}$, respectively.

The purpose of preconditioning is to reduce the number of iterations
and the computing time necessary to achieve a given accuracy. This means that
{\bf (a)} $V=V_1V_2$ has to be a sufficiently good approximation to
the inverse of $M$ 
and {\bf (b)} finding solutions for $V_1$ and $V_2$ should be
sufficiently cheap. 

Consider the decomposition
of $M$ into its diagonal, strictly lower $L$ and strictly upper $U$ triangular parts\
\[
M = I - L - U \, .
\]
The SSOR preconditioner is given by 
\begin{equation}
\label{SSOR_prec_eq}
V_1 = I - L, \enspace V_2 = I-U.
\end{equation}

For the SSOR preconditioner we have 
$V_1 + V_2 - M  = I$. This relation
can be exploited through the `Eisenstat-trick' \cite{Eis81}:
with $V_1^{-1}MV_2^{-1} = V_2^{-1} + V_1^{-1}(I-V_2^{-1})$, the
matrix vector product $w = V_1^{-1}MV_2^{-1}r$ can economically be
computed in the form
\[
v = V_2^{-1}r, \enspace u = V_1^{-1}(r-v), \enspace w = v + u.
\]

Note that multiplications with $M$ are completely
avoided in this formulation, the only matrix operations being
multiplied with are $I-L$ and $I-U$.
Since these matrices are triangular, the solutions can be computed 
directly via forward or backward substitution.
 
The preconditioned residuals 
$\tilde{r}_i$ are related to the unpreconditioned residuals \cite{Wupp}.
Upon successful stopping, one can compute $r_i$ and restart if the
solution is not yet accurate enough. 

\section{ORDERINGS}
In eq. \ref{wilson_eq} with the Wilson fermion
matrix $M$ we have the freedom to choose any ordering scheme for the lattice
points $x$. Different orderings yield different matrices $M$,
which are permutationally similar to each other.
%
%
\begin{figure}[htb]
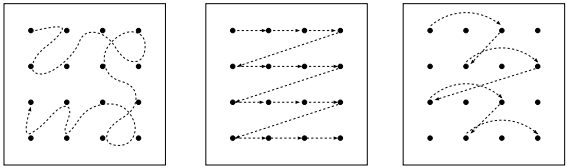

\smallpspicture{orderings.eps}
\caption[dummy]{\label{orderings} Random, natural and oe orderings of
the lattice sites defining three matrix multiplications
which are permutationally equivalent.}
\end{figure}
\vspace{-.4cm}
Consider an arbitrary ordering of the lattice points. For
a given grid point $x$, the corresponding row in the matrix $L$ or $U$
contains exactly the coupling coefficients of those nearest neighbors
of $x$ which have been numbered before or after $x$, respectively.  Therefore,
a generic formulation of the forward or backward solution for this ordering is
given by the rules
\begin{itemize}
\item
touch every site
\item forward solve: respect site numbered before
\item backward solve: respect site numbered after
\end{itemize}

In this context, the odd-even ordering, so far generally considered as the only
successful preconditioner in a parallel computing environment, 
is seen to be a specific example of SSOR preconditioning (in odd-even
ordering all odd lattice points are numbered before the even ones). 
In traditional QCD computations, the odd-even preconditioning is {\em
  not} implemented by using the above formulation of the forward (and
backward) solvers, as for this particular ordering the
inverses of $I-L$ and $I-U$ can be determined directly.

Defining $M$ with the natural ( global lexicographic [gl]) ordering  (fig.1) 
leads to a further improvement over odd-even preconditioning as far as
the number of iterations is concerned \cite{Oy85} .  However, its
parallel implementation turned out to be impractical \cite{HOCKNEY}. 

\section{PARALLELISATION}
Unlike the lexicographical ordering, the ordering we
propose now is adapted to the parallel computer used to solve eq. 
\ref{wilson_eq}.  We assume that the processors of the parallel
computer are connected as a $p_1 \times p_2 \times p_3 \times p_4$
4-dimensional grid. The space-time lattice can be matched to the processor
grid in a natural manner, producing a local lattice of size
$n^{loc}_1 \times n^{loc}_2 \times n^{loc}_3 \times n^{loc}_4$ with
$n^{loc}_i = n_i/p_i$ on each processor.

Let us partition the whole lattice into $n^{loc} = n^{loc}_1 n^{loc}_2
n^{loc}_3 n^{loc}_4$ groups.  Each group corresponds to a fixed
position of the local grid and contains all grid points appearing at
this position within their respective local grid.

\begin{figure}[htb]
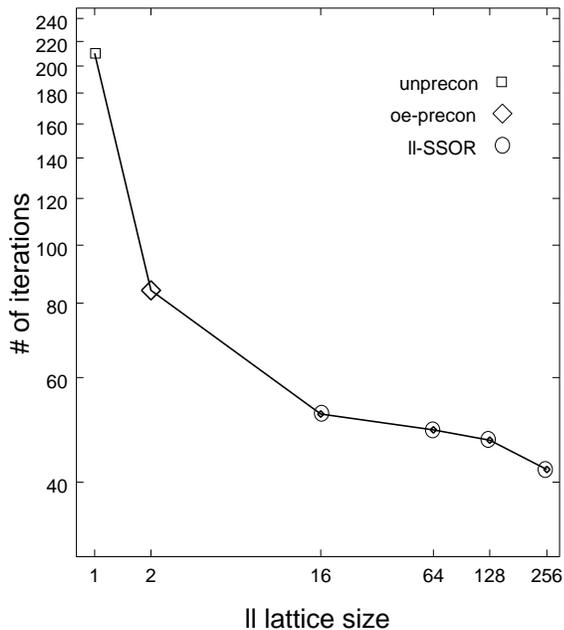

\smallpspicture{saturn.eps}
\caption[dummy]{\label{locvol}Local lexicographic ordering with 
different local volumes using BiCGStab. Reference point $( \Diamond ) $
refers to o-e preconditioning \cite{FHNLS94}.}
\end{figure}

We now consider a natural ordering on each local grid, which allows a
coherent update of corresponding sites on each processor in {\it parallel}.
Inter-node communication has to be done when the local grid border
is touched. Fig.\ref{locvol} shows the speedups of various local
lattices sizes 
($16=2^4, 64=2^2 \cdot 4^2, 128=2\cdot 4^3, 256=2\cdot 8 \cdot4^2$). It should be
noted that the number of floating point operations per iteration is the same in
each case.

\section{APPLICATION}
Our numerical tests of the locally lexicographic SSOR preconditioner
were performed on APE100/Quadrics machines (Q4,QH4), a SIMD parallel
architecture optimized for fast floating point arithmetic on block
data-structures like $3\times 3$ SU(3) matrices. We applied the $ll$
preconditioner to quark propagator calculations and chiral Hybrid
Monte Carlo simulations with Wilson fermions on large lattices. The
CPU time gains  for the computation of propagators are shown in fig.3.
In our current HMC implementation on a QH4 with a $24^3 \times 40 $
lattice close to the chiral regime we found a speedup in convergence
rate by a factor 2, corresponding to a gain in CPU time of 70\% in our
implementation.

\begin{figure}[htb]
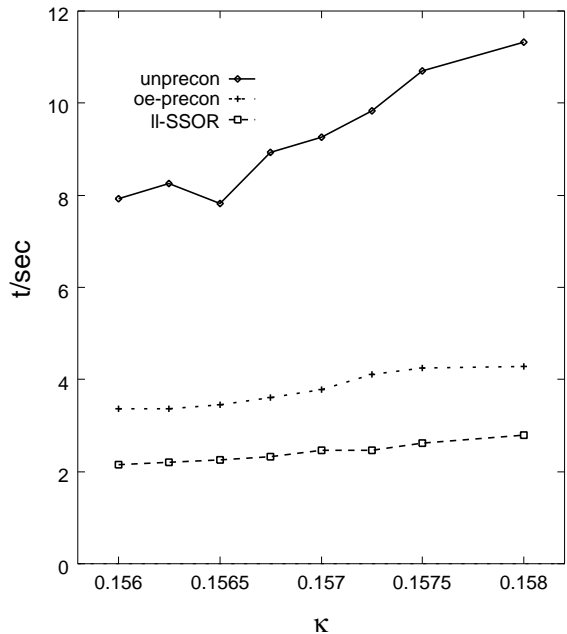

\smallpspicture{sea-val-time.eps}
\caption[dummy]{\label{sea-val}CPU times needed for the quark
propagator calculation on a $8^3 \times 16$ lattice at $\beta=5.6$ on
the Q4 with local lattice size $256=(4,4,8,2)$.}
\end{figure}

\section{CONCLUSION}
We have presented a new local grid point ordering scheme that allows
to carry out efficient preconditioning of Krylov subspace solvers.
The number of iterations as well as the required floating point
operations are reduced by a factor of 2 for local lattice sizes ($>$ 128 ).
This corresponds to an implementation and machine  dependent gain in CPU
time of 30\% - 70\%.

\end{document}